# Neuromorphic Metasurface

Zhicheng Wu, Ming Zhou, Erfan Khoram, Boyuan Liu, Zongfu Yu

Department of Electrical and Computer Engineering, University of Wisconsin, Madison, WI 53706

Abstract

**Metasurfaces have been used to realize optical functions such as focusing and beam steering. They use sub-wavelength nanostructures to control the local amplitude and phase of light. Here we show that such control could also enable a new function of artificial neural inference. We demonstrate that metasurfaces can directly recognize objects by focusing light from an object to different spatial locations that correspond to the class of the object.**

Optical neuromorphic computing offers an alternative approach to realize artificial neural computing. It has several potential advantages compared to digital neural computing such as the ultra-fast speed and ultra-low energy consumption. Several architectures have been demonstrated based on integrated silicon photonics [1], diffractive optics [2], and nanophotonic random structure [3]. In this paper, we introduce another platform to realize artificial neural computing based on metasurfaces. Metasurfaces were developed to perform arbitrary phase front engineering[4]. Their optical functions are realized by the resonant scattering of arrays of nanoscale scatterers fabricated on a flat surface. It is compatible with today's nanofabrication and can be mass-produced at low cost [5]. Here, we use these nanoscale scatterers to perform neural computing. It leverages the platform of flat optics to realize high-density integration. We describe the design procedures and demonstrate direct image recognition of handwritten digits.

The concept is illustrated in Fig. 1. An object, such as a handwritten digit, is illuminated by a plane wave. The scattered light is then processed by a multi-layer neuromorphic metasurface, which consists of arrays of nanoribbons. By changing the size of the ribbons, we can control the amplitude and the phase of scattered light as shown in Fig. 1, which lead to strong interference of light waves passing through the metasurface. With a few layers of metasurfaces, the output light becomes a focused beam and is directed toward a spatial location corresponding to the value of the handwritten digit. The widths of the nanoribbons are the trainable parameters, which are learned through a training process similar to stochastic adjoint optimization [3].

This work is related to the diffractive neural network demonstrated by Xing Lin in 2018 [2], where they use the thickness of the material that light passes through to modulate the phase. Varying thicknesses is not easily compatible with nanofabrication for large scale integration. By using metasurfaces, we can tune the phase delay using the lateral dimension so that the device can be made easily with today's lithography. In order to account for the phase delay caused by lateral structures, full-wave electromagnetic modeling must be used. Such full-wave modeling can be extremely expensive. Here we describe the approaches to reduce the computational load.  Also related to the work is [3], where continuous media is used for neural computing. Here metasurface can be fabricated on flat surfaces, greatly simplifying the fabrication process.

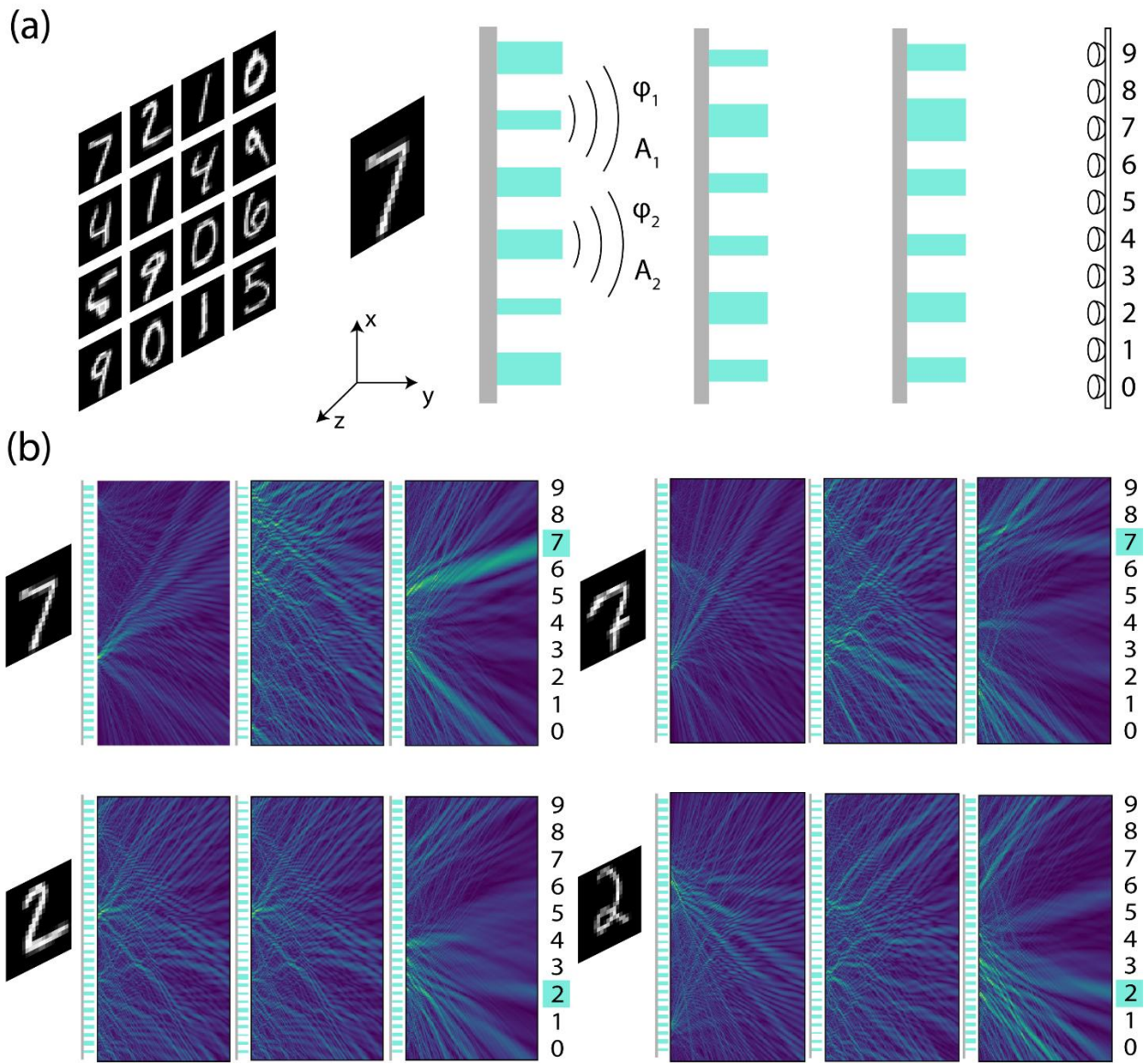

Fig. 1. (a) Schematic of the neuromorphic metasurface. The neuromorphic metasurface consists of multiple layers of nanostructures, which are composed of an array of nanoribbons on top of a dielectric substrate. A handwritten digit is illuminated by a plane wave and the scattered light then is processed by the neuromorphic metasurface. By changing the sizes of ribbons, the phase and amplitude of the transmitted light after each layer can be modified. After multiple layers, the transmitted field can be focused to specific photodetectors, which are labeled by the values of the handwritten digits, i.e. 0 to 9. (b) Intensity distribution of the transmitted light after each layer in a 3-layer neuromorphic metasurface. Handwritten digits of 7 and 2 with different writing style are used as examples. Despite the different writing styles, the transmitted light is always focused to the spot corresponding to the value of the handwritten digit. Here, we normalize the intensity of the transmitted light after each layer to its maximum for clarity.

We use a specific example to illustrate how to design neuromorphic metasurfaces. The goal is to recognize hand-written digits such as the one shown in Fig. 1. We use the database MNIST [6], which contains 60000 different handwritten digits. We use 50000 examples for training and 10000 examples for the test. The neuromorphic metasurface should correctly recognize the value of the digits despite their different handwriting styles. We divide the database into two groups. The first group, the training set, is used to train the metasurface. The second group, the test set, is used to test the utility of metasurface. A plane wave illuminates on the handwritten digits and then passes through the metasurface which scatters the light in a way that is equivalent to artificial neural computing. The output light will focus on one of ten different spatial locations that correspond to different values of digits. Here below, we will use two-dimensional metasurfaces to illustrate the design process. The three-dimensional design follows the same procedure.

The metasurface consists of a large area of subwavelength scattering elements. Full-wave tools such as finite-difference time-domain are too computationally expensive for this type of multiscale problem. To obtain the full-wave electromagnetic properties without losing speed, we use locally periodic approximation [7]–[16]. It assumes the metasurface is locally periodic: the transmitted field in any small region is approximately the same as the transmitting from a periodic surface. The field amplitude and phase immediately after a scattering element are calculated by a full-wave simulation assuming a periodic boundary condition, as shown in Figure 2.

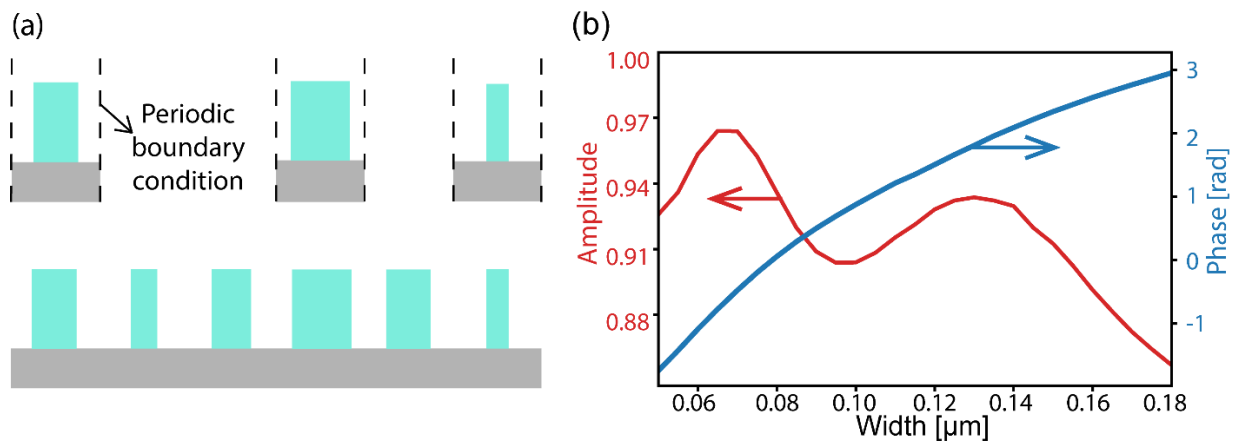

Fig. 2. (a) Schematic of locally periodic approximation. The metasurface consists of an array of $TiO_2$ pillars on top of a $SiO_2$ substrate. For plane-wave incidence from the bottom side of the substrate, we set up a periodic boundary condition around each pillar. The local field of the transmitted light above each pillar then is approximated by that of the corresponding periodic array. (b): The phase (blue) and amplitude (red) of transmitted light as a function of the width of the pillar under normal plane-wave incidence. The results are obtained from a full-wave simulation of a periodic array of pillars, which only takes a few minutes.

By using a small full-wave simulation to obtain the local field for each element, we can assemble the field along the plane right after the metasurface. Then we can use near-to-far field transformation to calculate far-field distribution. Compared to the Rayleigh-Sommerfeld diffraction equation used in [2], the local periodic approximation takes into account the wave effect of structured scatterers. Compared to the finite-difference full-wave method used in [3], this method is much faster. The comparison of this method with rigorous modeling can be found in [7].

Here we use TiO2 pillars on a SiO2 substrate to construct the metasurface [13]. The thickness of the substrate is 300nm. The height of the pillar is fixed to 600nm and the pitch is fixed to 235nm. We vary the pillar's width from 50 nm to 180 nm. The phase $\varphi_n(w)$ and amplitude $A_n(w)$ of the transmitted light is shown in Fig. 3, where $w$ is the width of the pillar and the learnable parameter, and the subscript $n$ represents the normal incident direction. The operating wavelength is 700 nm.

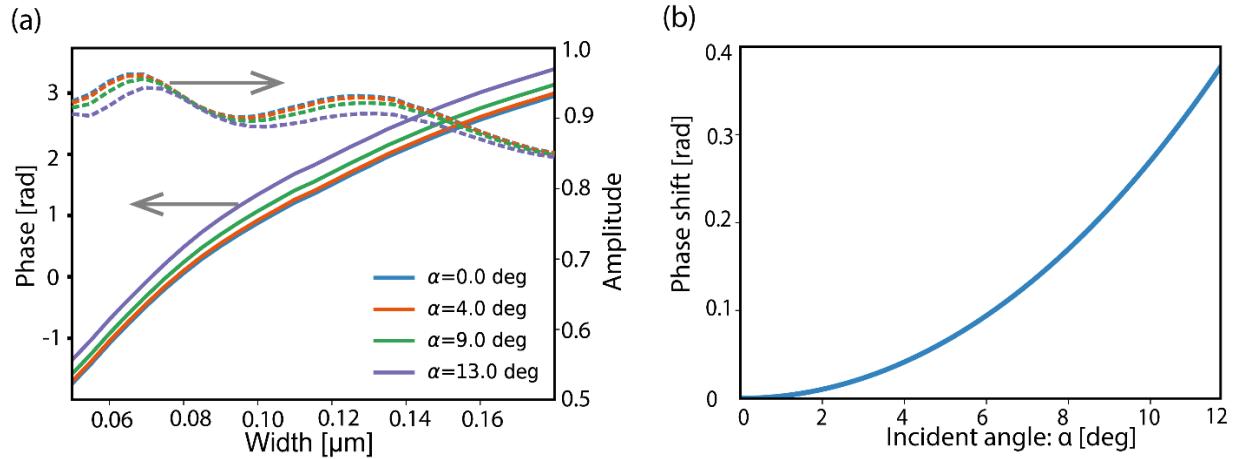

Fig. 3. (a) The phase (solid lines) and amplitude (dashed lines) of the transmitted field as a function of the width of the pillar for the different incident angle. The incident wavelength is 700 nm. The phase response curve shifts upwards as we increase the incident angle, while the amplitude response does not vary significantly with the incident angle. (b): The shift of the phase response as a function of incident angle. The shift increases nonlinearly with increasing incident angle.

The input wavefront to neuromorphic metasurfaces is generally much more complex than plane waves. Since we have to use a plane wave as the incident condition when applying the locally periodic approximation, we first decompose the incoming wavefront $E(x)$ using Fourier basis $E_k = \sum_x E(x)e^{ikx}$ and then simulate the response of metasurface under each individual plane waves $E_k$. Then we sum all the contribution of plan waves together. We could also safely neglect plane waves with large wave vector $k$ because of the large distances between different metasurfaces and between the object and the metasurface.

The phase delay and amplitude modulation change for plane waves incident from different angles. Figure 3 shows the response of the pillars for different incident angles. The phase response curve shifts horizontally but the amplitude does not vary significantly. This observation allows us to further accelerate the computation by approximating the angular response with $E_c(x) = \sum_k E_k e^{-ikx+i\theta_k}$. The phase compensation $\theta_k$ accounts for the difference of phase delay compared to the normal incident wave $k = 0$ (Fig. 3b). Now we can calculate the scattering field using transmission of normally incident plane wave with the corrected wavefront compensated for the different incident angle. The transmitted wave is calculated by the convolution $E_c(x) * A_n(w(x))e^{i\varphi_n(w(x))}$ where $w(x)$ is pillar width at position x.

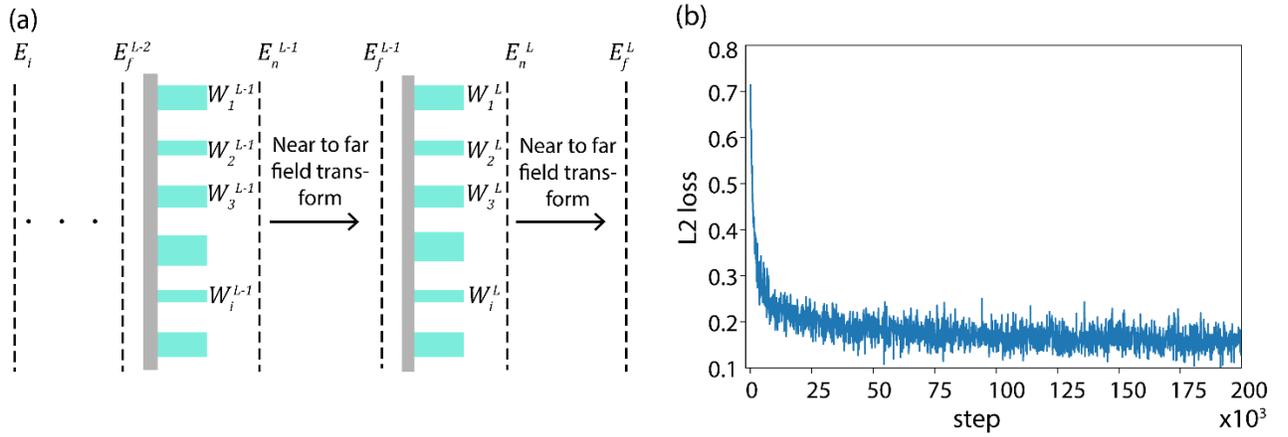

Fig4. (a) Forward propagation in a neuromorphic metasurface that has $L$ layers. At the $l^{th}$ layer, the pillars take the far-field from the $(l-1)^{th}$ layer $E_f^{l-1}$ as input and process it. The output near field $E_n^l$ is obtained by using locally periodic approximation. The corresponding far-field $E_f^l$ at the $(l+1)^{th}$ layer then is obtained by using near to far-field transformation. The final output of the neuromorphic metasurface is the intensity of light $|E_f^L|^2$. (b) The training loss of a 5-layer neuromorphic metasurface as a function of training steps. Each step, we use 1 training sample to update widths of the pillars. The training dataset is reshuffled every 50000 steps.

We now discuss the training process. The output of the neuromorphic metasurface is defined by the distribution of electric field intensity on a plane behind the last layer of the metasurface. In the 2D case, the output is $y(x) = |E_f^L|^2$.

Here we use subscript $f$ to indicate the far-field distribution of light after passing the $l_{th}$ metasurface layer. The training target for the output is $y_t(x) = I_0 \exp(-\frac{(x-x_t)^2}{2*\sigma^2})$, where $t$ is the value of the handwritten digits. $x_t$ are the locations where we would like output light to focus on. Locations for different digit value are evenly distributed on the output plane. One can also choose other training target as long as it serves the purpose of classification. In our 2-D case, the peak position of the target intensity $x_t$ for different digit number are equally spaced by 9.4um, and the variance $\sigma$ of the target intensity is 2.35um.

Training the metasurface is a gradient descent process that minimizes the loss function $L$. Here we use the L2 distance between the metasurface output and the target output $L = \sqrt{(y(x) - y_t(x))^2}$. Unlike typical optimization used in nanophotonics and metasurfaces [7], [17], the gradient descent used here is stochastic.

Next, we discuss how to compute $L$ and its gradients. First, we try to formulate the relation between the far-field outputs of the $l^{th}$ layer and the $(l-1)^{th}$ layer. This relation depends on the width of pillars $w_i^l$ in the $l^{th}$ layer. The far-field output is calculated from the near-field $E_n^l(x)$ through a near-to-far field transformation[18]:

$$E_f^l(x) = -\int_{surface} G(x, x')E_n^l(x')dx'$$

where $G$ is a Hankel function $G(x,x') = -\frac{ik}{4}H_1^{(1)}(kr)\hat{n} \cdot \frac{r}{r}$. Here $k = \frac{2\pi}{\lambda}$, $r = x - x'$, and $r = |r|$. The near field is obtained through local periodic approximation $E_n^l(x) = \sum_k E_k e^{-ikx+i\theta_k} * A_n(w^l(x))e^{i\varphi_n(w^l(x))}$. Here $E_k$ is the Fourier component of $E_f^{l-1}$, the far-field output of the $(l-1)_{th}$ layer. This series of calculation that connects $E_f^l(x)$ and $E_f^{l-1}(x)$ can all be represented as matrix operation and implemented in Tensorflow[19]. For example, the integral is changed to summation and can be expressed as a matrix multiplication: $\boldsymbol{E_f} = \boldsymbol{G} \cdot \boldsymbol{E_n}$, where $G_{ij} = G(x_i, x_j')$, $\boldsymbol{E_n}(j) = E_n(x_j')$, and $\boldsymbol{E_f}(i) = E_f(x_i)$. We neglect the reflection of the metasurfaces because the low index substrate used here results in weak reflection.

We now are ready to calculate the derivative of the loss function with respect to the pillar widths $\frac{\partial L}{\partial w(x)}$. The calculation can be divided into two steps because $\frac{\partial L}{\partial w(x)} = \frac{\partial L}{\partial E_n^l(x)} \frac{\partial E_n^l(x)}{\partial w_l(x)}$. The first term is the derivative of the loss function with respect to each layer's output near field, which is calculated by following the chain rule of derivative $\frac{\partial L}{\partial y} \frac{\partial y}{\partial E_n^L} \frac{\partial E_n^L}{\partial E_n^{L-1}} \cdots \frac{\partial E_n^{l+1}}{\partial E_n^l(x)}$ in Tensorflow. The second is the derivative of the output field with respect to the pillar widths $\frac{\partial E_n^l(x)}{\partial w_l(x)} = \frac{\partial E_n^l(x)}{\partial \varphi(x)} \frac{\partial \varphi(x)}{\partial w_l(x)} + \frac{\partial E_n^l(x)}{\partial A(x)} \frac{\partial A(x)}{\partial w_l(x)}$. The phase $\varphi(w^l(x))$ and amplitude $A(w^l(x))$ as a function of pillar width are shown in Figure 3, which allows us to easily calculate $\frac{\partial \varphi}{\partial w}$ and $\frac{\partial A}{\partial w}$. One difference from the conventional deep learning is that the learnable parameters are also constrained by the physical limit of pillar sizes.

Generally, the input of neuromorphic metasurface is the light scattered by the object. In the simulation, the input is replaced by the image of the object. For the 2-D case, we also vectorize the image of the handwritten digit number. The original image is resized to 20 by 20 pixels and converted to a 1 by 400 vector and the intensity is normalized from 0 to 1. Then, we can set the intensity of the vectorized image as the amplitude of the input field. The phase of the input field is set to be the same. The input field is polarized in the z-direction such that field can be treated as a complex scalar in simulation. The wavelength is 700 nm. At this frequency, the response of periodic TiO2 structure changes smoothly when the width of pillar changes. To match the size of the input vector, each layer of neuromorphic metasurface also contains 400 elements. The pitch is 235 nm wide, and the total length of the metasurface is 94 m. The distance between the two adjacent layers is 188 m.

The training process of the 5-layer neuromorphic metasurface is shown in Figure 4 (b). Each data point is the averaged L2 loss over 100 training samples. The computation took 13 hours on an Intel Core i5-4430 CPU 3.00GHZ×4.

| Layer | 2 | 3 | 4 | 5 | 6 |
|---|---|---|---|---|---|
| Accuracy | 80% | 85% | 88% | 89% | 90% |

Table1: Accuracies of MNN for different number of layers

The neuromorphic metasurface starts to show its utility even with just two layers of metasurfaces. It can achieve 80% accuracy for MNIST classification. It means that 8 out 10 times, this double-layer metasurface can focus the light on the right location based on the meaning of the handwritten digit. It is a remarkable focusing effect compared to traditional metasurfaces that focus all light to a single spot. The accuracy can further improve when more layers are used. These results are shown in Table 1. Figure 5 shows the light field propagation in a 5-layer neuromorphic metasurface before and after training. It can be seen that at the beginning of the training, light is directed to a random distribution. After training, light is focused on the right classification spot.

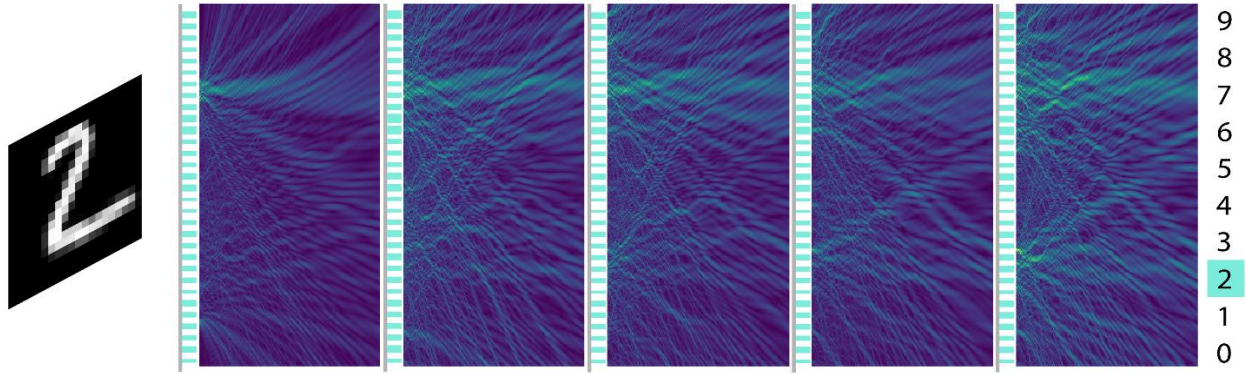

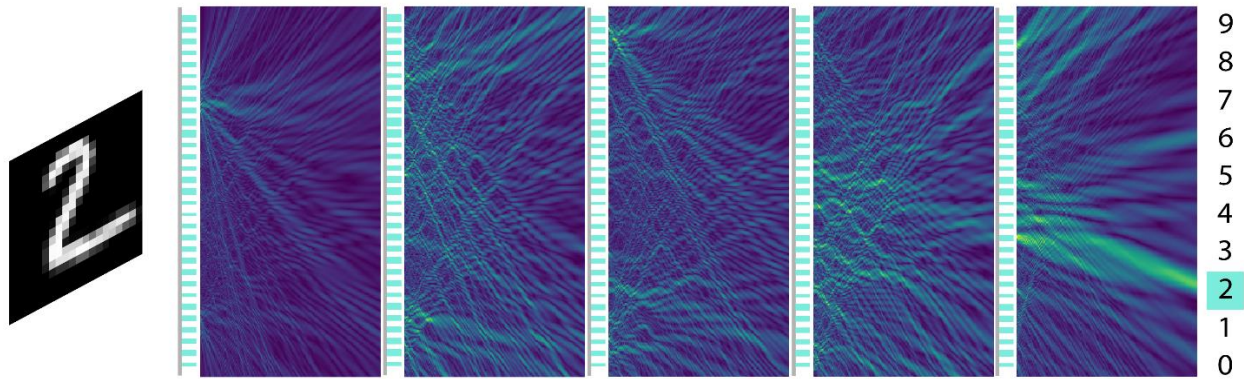

Fig. 5 Comparison of light propagation before (a) and after (b) training for a 5-layer neuromorphic metasurface. The input object is a handwritten digit of 2. Before training, the widths of the pillars are randomly initialized and the transmitted light is randomly distributed at the detector. After training, the transmitted light is directed to detector 2, which corresponds to the input handwritten digit. Here, the intensity distribution of the transmitted light after each layer is also normalized by its maximum for clarity.

Unlike our previous work demonstrated in [3], here we did not use nonlinear activation. In this simple recognition task, nonlinear activation does not significantly enhance performance. But nonlinear activation is crucial for more complex tasks such as face recognition. Nonlinear materials such as a layer of saturable absorber can be easily fabricated into multi-layer metasurfaces. In [3], we solve nonlinear full-wave Maxwell's equation to account for nonlinear activation. To make the computation more manageable, here we did not apply nonlinear activation for this multi-scale metasurfaces. Further work is needed to significantly speed up the electromagnetic modeling of nonlinear materials to be used for metasurfaces.

Acknowledgment: The work was partially supported by DARPA under a YFA program and NSF under ECCS 1641006

[1] Y. Shen *et al.*, "Deep learning with coherent nanophotonic circuits," *Nature Photonics*, vol. 11, no. 7, pp. 441–446, Jul. 2017.
[2] X. Lin *et al.*, "All-optical machine learning using diffractive deep neural networks," *Science*, vol. 361, no. 6406, pp. 1004–1008, Sep. 2018.
[3] E. Khoram *et al.*, "Nanophotonic media for artificial neural inference," *Photon. Res., PRJ*, vol. 7, no. 8, pp. 823–827, Aug. 2019.
[4] N. Yu *et al.*, "Light Propagation with Phase Discontinuities: Generalized Laws of Reflection and Refraction," *Science*, vol. 334, no. 6054, pp. 333–337, Oct. 2011.
[5] N. Yu and F. Capasso, "Flat optics with designer metasurfaces," *Nature Materials*, vol. 13, no. 2, pp. 139–150, Feb. 2014.
[6] "MNIST handwritten digit database, Yann LeCun, Corinna Cortes and Chris Burges." [Online]. Available: http://yann.lecun.com/exdb/mnist/. [Accessed: 12-Jul-2019].
[7] R. Pestourie, C. Pérez-Arancibia, Z. Lin, W. Shin, F. Capasso, and S. G. Johnson, "Inverse design of large-area metasurfaces," *Opt. Express, OE*, vol. 26, no. 26, pp. 33732–33747, Dec. 2018.
[8] F. Aieta *et al.*, "Aberration-Free Ultrathin Flat Lenses and Axicons at Telecom Wavelengths Based on Plasmonic Metasurfaces," *Nano Lett.*, vol. 12, no. 9, pp. 4932–4936, Sep. 2012.
[9] A. Arbabi, E. Arbabi, Y. Horie, S. M. Kamali, and A. Faraon, "Planar metasurface retroreflector," *Nature Photonics*, vol. 11, no. 7, pp. 415–420, Jul. 2017.
[10] M. Khorasaninejad *et al.*, "Visible Wavelength Planar Metalenses Based on Titanium Dioxide," *IEEE Journal of Selected Topics in Quantum Electronics*, vol. 23, no. 3, pp. 43–58, May 2017.
[11] F. Aieta, M. A. Kats, P. Genevet, and F. Capasso, "Multiwavelength achromatic metasurfaces by dispersive phase compensation," *Science*, vol. 347, no. 6228, pp. 1342–1345, Mar. 2015.
[12] M. Khorasaninejad *et al.*, "Achromatic Metasurface Lens at Telecommunication Wavelengths," *Nano Lett.*, vol. 15, no. 8, pp. 5358–5362, Aug. 2015.
[13] M. Khorasaninejad *et al.*, "Achromatic Metalens over 60 nm Bandwidth in the Visible and Metalens with Reverse Chromatic Dispersion," *Nano Lett.*, vol. 17, no. 3, pp. 1819–1824, Mar. 2017.
[14] "OSA | Controlling the sign of chromatic dispersion in diffractive optics with dielectric metasurfaces." [Online]. Available: https://www.osapublishing.org/optica/abstract.cfm?uri=optica-4-6-625. [Accessed: 15-Jul-2019].
[15] V.-C. Su, C. H. Chu, G. Sun, and D. P. Tsai, "Advances in optical metasurfaces: fabrication and applications [Invited]," *Opt. Express, OE*, vol. 26, no. 10, pp. 13148–13182, May 2018.

[16] B. Groever, C. Roques-Carmes, S. J. Byrnes, and F. Capasso, "Substrate aberration and correction for meta-lens imaging: an analytical approach," *Appl. Opt., AO*, vol. 57, no. 12, pp. 2973–2980, Apr. 2018.

[17] A. Y. Piggott, J. Lu, K. G. Lagoudakis, J. Petykiewicz, T. M. Babinec, and J. Vučković, "Inverse design and demonstration of a compact and broadband on-chip wavelength demultiplexer," *Nature Photonics*, vol. 9, no. 6, pp. 374–377, Jun. 2015.

[18] A. Taflove and S. C. Hagness, *Computational electrodynamics: the finite-difference time-domain method*, 3rd ed. Boston, MA: Artech House, 2005.

[19] "TensorFlow," *TensorFlow*. [Online]. Available: https://www.tensorflow.org/. [Accessed: 12-Jul-2019].